\documentclass[aps,twocolumn,showpacs,preprintnumbers,amsmath,amssymb,superscriptaddress,longbibliography]{revtex4-2}
\usepackage{graphicx}
\usepackage{dcolumn}
\usepackage{bm}
\usepackage{braket}
\usepackage{amsmath}
\usepackage[normalem]{ulem}

\begin{document}

\title{Detecting Topological Phases Through Dynamics of Chiral Density in Momentum-space Lattices}

\author{Rui Tian}
\affiliation{Ministry of Education Key Laboratory for Nonequilibrium Synthesis and Modulation of Condensed Matter,Shaanxi Province Key Laboratory of Quantum Information and Quantum Optoelectronic Devices, School of Physics, Xi'an Jiaotong University, Xi'an 710049, China}

\author{Xiayao He}
\affiliation{Ministry of Education Key Laboratory for Nonequilibrium Synthesis and Modulation of Condensed Matter,Shaanxi Province Key Laboratory of Quantum Information and Quantum Optoelectronic Devices, School of Physics, Xi'an Jiaotong University, Xi'an 710049, China}

\author{Tianhao Wu}
\affiliation{Ministry of Education Key Laboratory for Nonequilibrium Synthesis and Modulation of Condensed Matter,Shaanxi Province Key Laboratory of Quantum Information and Quantum Optoelectronic Devices, School of Physics, Xi'an Jiaotong University, Xi'an 710049, China}

\author{Dongdong Pu}
\affiliation{Ministry of Education Key Laboratory for Nonequilibrium Synthesis and Modulation of Condensed Matter,Shaanxi Province Key Laboratory of Quantum Information and Quantum Optoelectronic Devices, School of Physics, Xi'an Jiaotong University, Xi'an 710049, China}

\author{Shuai Li}
\affiliation{Ministry of Education Key Laboratory for Nonequilibrium Synthesis and Modulation of Condensed Matter,Shaanxi Province Key Laboratory of Quantum Information and Quantum Optoelectronic Devices, School of Physics, Xi'an Jiaotong University, Xi'an 710049, China}

\author{Bo Liu}
\email{liubophy@gmail.com}
\affiliation{Ministry of Education Key Laboratory for Nonequilibrium Synthesis and Modulation of Condensed Matter,Shaanxi Province Key Laboratory of Quantum Information and Quantum Optoelectronic Devices, School of Physics, Xi'an Jiaotong University, Xi'an 710049, China}

\begin{abstract}
    Momentum-space lattices offer a versatile platform for quantum simulation in particular emulating topological states of matter.
	In this work, we propose a spectroscopic protocol via utilizing the dynamics of the chiral operator to detect the topological properties of a momentum-space Su-Schrieffer-Heeger chain. We show that the time evolution of the  chiral density is directly linked to the local density of state (LDOS). By performing a Fourier transformation of that, we reconstruct both the LDOS and density of states of the system, enabling the unambiguous signature of zero-energy edge modes and the bulk-boundary correspondence. This dynamical approach provides a diagnostic tool for exploring topological phases in synthetic quantum matter.
\end{abstract}

\maketitle

\section{Introduction}
\label{introduction}

The realization of topological phases in highly controllable quantum systems has become a central theme in modern condensed matter physics and quantum simulation \cite{qi2011topological,senthil2015symmetry,wen2017colloquium,hasan2010colloquium,ren2016topological}. Among various experimental platforms, ultracold atoms in optical lattices stand out as a premier candidate for exploring exotic physics due to their unprecedented level of precise control \cite{zhang2018topological,schafer2020tools,gross2017quantum,bloch2005ultracold,bloch2012quantum,goldman2016topological}. 
In recent years, the development of synthetic dimensions has significantly expanded the scope of quantum simulation \cite{ozawa2019topological,lustig2019photonic,lustig2021topological}, one of the prototype examples is the momentum-space lattice, where discrete momentum of atoms serves as the lattice sites \cite{gadway2015atom}. Unlike the traditional real-space lattices, they're constructed in momentum-space and are intrinsically immune to disorder \cite{meier2016observation,an2018correlated,agrawal2024two,an2021nonlinear,lapp2019engineering,xie2019topological,an2017direct,meier2018observation,yu2025comprehensive}. Furthermore, by utilizing multi-frequency laser engineering scheme to drive momentum transferring, different momentum lattices can be designed. It makes momentum lattices being an ideal platform for exploring various lattice model, such as the Su-Schrieffer-Heeger (SSH) chain \cite{meier2016observation}.

It is known that one of the remarkable properties of topological phase is the existence of the topological edge states.
The local density of states (LDOS) is one of the fundamental quantities that can reveal the energy-resolved spatial distribution and can be used as a "smoking gun signature" for identifying topological boundary modes \cite{peterson2020fractional,grinberg2020trapped,peterson2021trapped,liu2021bulk,yamada2022bound,liang2023fractional,ge2023direct,zheng2023fractional,zhang2023observations}. 
As distinct from the real-space system, in synthetic momentum space, where sites are defined in velocity space rather than coordinate space, an analogue for such a local probe is still missing. 
In this work, we introduce a spectroscopic protocol that leverages chiral operator dynamics \cite{zhang2024dynamical} to uncover the topological spectral density in momentum-space lattices. 
By taking a momentum-space SSH chain as an example, we show that the dynamics of chiral density can unambiguously resolve sharp zero-energy LDOS peaks, indicating the existence of topological edge modes at the synthetic lattice boundaries.

\begin{figure*}
	\centering
	\includegraphics[width=1\textwidth]{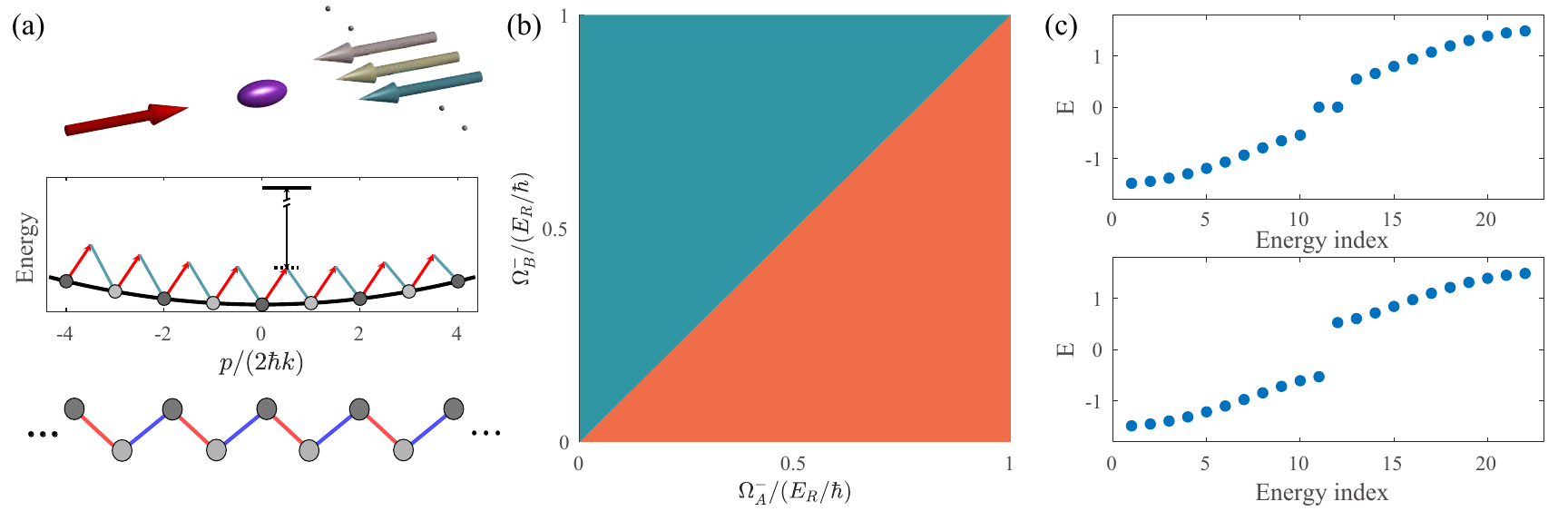}
	\caption{
		(a) Scheme for realizing momentum-space lattices in atomic gases. Atoms are driven by counter-propagating laser fields and the multi-frequency left-propagating beam is tailored with specific frequency components to engineer alternating tunneling amplitudes $v$ and $w$ between discrete momentum states.
		(b) Phase diagram of the momentum-space SSH model as a function of Rabi couplings $\Omega_A^-$ and $\Omega_B^-$. Different colors mark distinct topological phases, i.e. topological nontrivial (blue) and trivial (orange) regimes.
		(c) Open-boundary-condition (OBC) energy spectrum shows the emergence (absence) of topological zero-energy modes for the topological nontrivial (trivial) case.
	}
	\label{FIG:fig1}
\end{figure*}

\section{Momentum lattice in atomic gases}
We consider an ensemble of two-level atoms with mass $M$, characterized by a ground state $|g\rangle$ and an excited state $|e\rangle$ separated by an energy $\hbar \omega_{eg}$.The single-particle Hamiltonian for atoms interacting with a driving laser field $\mathbf{E}(\mathbf{x}, t)$ is given by
\begin{equation}
	\hat{H} = \frac{\hat{\mathbf{p}}^2}{2M} + \hbar \omega_e |e \rangle \langle e | + \hbar \omega_g |g \rangle \langle g | - \mathbf{d}\cdot \mathbf{E}(\mathbf{x}, t) \ ,\label{EQ:Ham_lab}
\end{equation}
where $\hat{\mathbf{p}}$ captures the atomic momentum and $\mathbf{d}$ is the electric dipole. As illustrated in Fig.\ref{FIG:fig1}(a), the driving field $\mathbf{E}(\mathbf{x}, t)$ consists of a monochromatic right-propagating beam, $\mathbf{E}^+(\mathbf{x},t) = \mathbf{E}^+ \cos (\mathbf{k}^+ \cdot \mathbf{x} - \omega^+ t )$, and a multi-frequency left-propagating "control beam", $\mathbf{E}^-(\mathbf{x},t) = \sum_{j} \mathbf{E}^{-}_{j} \cos (\mathbf{k}^{-}_{j} \cdot \mathbf{x} - \omega^{-}_{j} t )$. We set the wavevectors as $\mathbf{k}^+ = k \hat{x}$ and $\mathbf{k}^{-}_{j} \simeq -k \hat{x}$, where $k=2\pi/\lambda$ corresponds to the laser wavelength $\lambda$. All frequency components are far-detuned from the atomic resonance by $\Delta \equiv \omega_{eg} - \omega^+ \simeq \omega_{eg} - \omega^{-}_{j}$, ensuring that the excited state $|e\rangle$ can be adiabatically eliminated.By utilizing a Bose-Einstein condensate (BEC) with a narrow momentum spread, we define a discrete basis of plane-wave momentum states $\{|n\rangle\}$, where each state carries a momentum $p_n = 2n\hbar k$ along the $\hat{x}$-direction. The interference between $\mathbf{E}^+$ and each component of $\mathbf{E}^-_j$ drives two-photon Bragg transitions, coupling adjacent momentum states $|n\rangle$ and $|n+1\rangle$. Under the weak-driving condition $\hbar \Omega_{j} \ll 8 E_R$ ($E_R = \hbar^2 k^2 / 2M$), the system effectively maps onto a tight-binding lattice in the momentum space \cite{gadway2015atom}. By precisely tailoring the frequency offsets $\omega^{-}_j$ and assigning alternating Rabi couplings $\Omega_{2n}^- \equiv \Omega_A^-$ and $\Omega_{2n+1}^- \equiv \Omega_B^-$, the one-dimensional Su-Schrieffer-Heeger (SSH) chain can be engineered \cite{meier2016observation}
\begin{equation}
	\hat{H}_{\mathrm{eff}} \approx \sum_n (v \hat{a}^\dag_{n} \hat{b}_n + w \hat{a}^\dag_{n+1} \hat{b}_n) + \mathrm{H.c.} \ ,\label{EQ:SSH_Ham}
\end{equation}
where $\hat{a}_n$ and $\hat{b}_n$ denote the annihilation operators for the two sublattices (even and odd momentum sites) within the $n$-th unit cell. The intracell and intercell tunneling amplitudes are given by $v=\Omega^{\ast -}_A \Omega^+/2 \Delta$ and $w=\Omega^{\ast -}_B \Omega^+/2 \Delta$, respectively.

\section{SPECTROSCOPIC DETECTION VIA CHIRAL DENSITY DYNAMICS}
A key advantage of the momentum-space lattices is the ability to perform site-resolved detection. To probe the local density of states (LDOS) and identify topological boundary modes, we employ a dynamical protocol based on the time evolution of the chiral operator.
In the momentum-space SSH chain,
each lattice site corresponds to a momentum state $|n\rangle$. 
The state initialization can be achieved through a sequence of pulsed Bragg diffractions, which can coherently transfer the BEC from the zero-momentum state to any target state $|n_0\rangle$ \cite{kozuma1999coherent}. 
To detect topological edge modes, we consider the system is initialized at a single site $|\bar{n}\rangle$ (e.g., at the boundary or in the bulk), which serves as an eigenstate of the chiral operator $\Gamma$ with eigenvalue $\pm 1$. The expectation value of the chiral operator, $\langle \Gamma(t) \rangle_{\bar{n}}$, is the central observable in our protocol. In the momentum-space lattice, this quantity is directly accessible through the standard absorption imaging following time-of-flight (TOF) expansion \cite{meier2016observation}. Since different momentum states $|n\rangle$ correspond to distinct velocities, they are spatially separated during TOF. The intensity of each cloud is proportional to the atomic population $P_{\bar{n}}(t) = |\langle \bar{n} | e^{-iHt}|\bar{n} \rangle|^2$ at that specific site. Consequently, the chiral operator expectation value can be computed as the difference in total populations between the two sublattices
\begin{equation}
	\langle \Gamma(t) \rangle_{\bar{n}} = \sum_{\bar{n} \in \mathrm{even}} P_{\bar{n}}(t) - \sum_{\bar{n} \in \text{odd}} P_{\bar{n}}(t) \ .
\end{equation}
By recording the evolution of $\langle \Gamma(t) \rangle_{\bar{n}}$ over a sufficiently long time $t$ and performing a Fourier transform, we can reconstruct the spectral density at site $\bar{n}$. 

\begin{figure*}
	\includegraphics[width=0.97\textwidth]{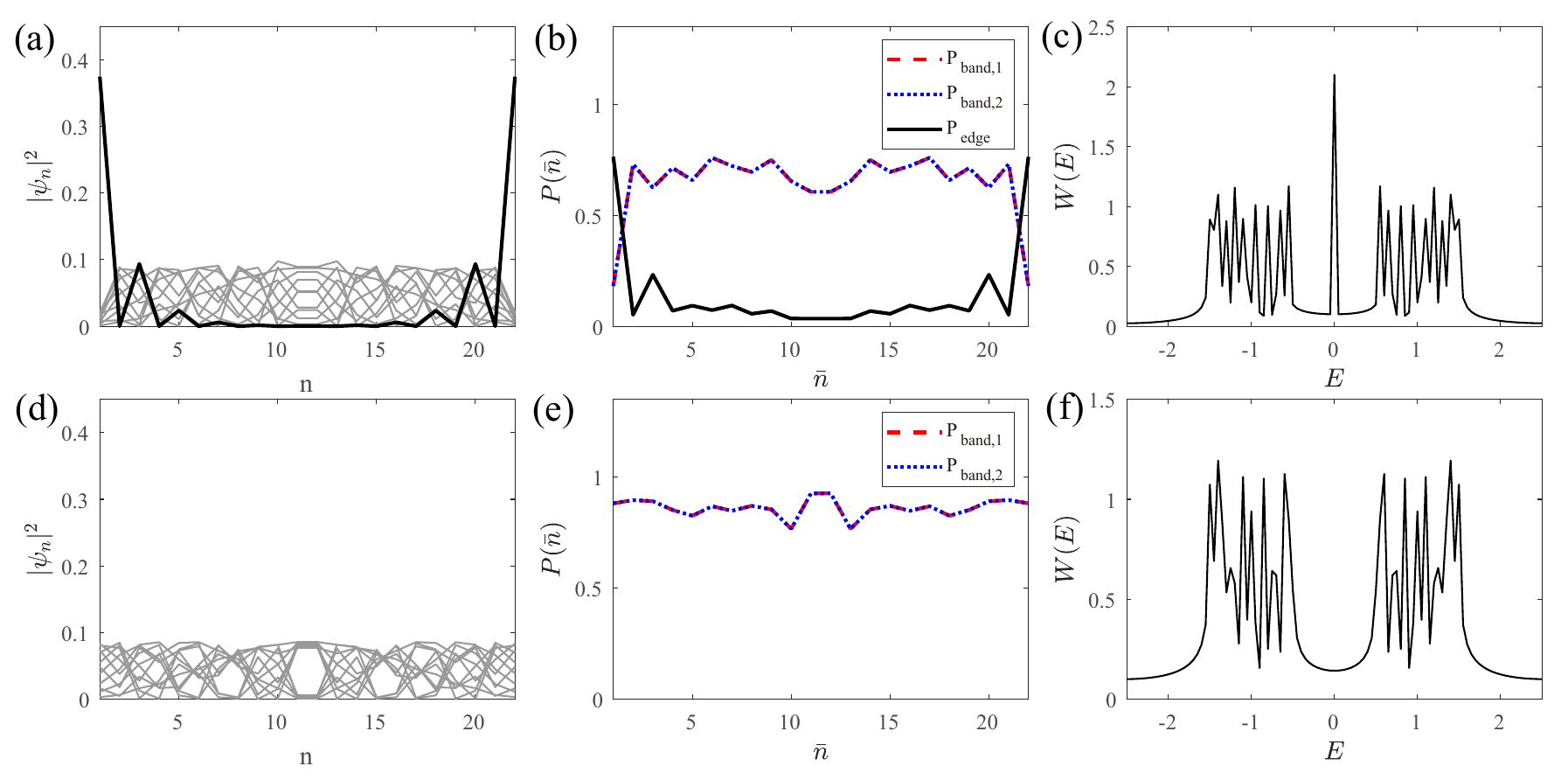}
	\caption{
		Left column: Spatial distribution of eigenstates $|\psi(n)|^2$ for (a) the topological non-trivial SSH chain ($v=0.5, w=1.0$) and (d) the trivial case ($v=1.0, w=0.5$). 
		Middle column: Integrated spatial weight $P(\bar{n})$: bulk states ($P_{\text{band,1}}$, $P_{\text{band,2}}$) and edge states ($P_{\text{edge}}$). In the topological phase (b), $P_{\text{edge}}$ is sharply peaked at the lattice boundaries, whereas in the trivial phase (e), only bulk state distributions are observed.
		Right column: $W(E)$ reconstructed via Fourier transformation of the chiral dynamics. The sharp peak at $E=0$ in the topological case (c) serves as the definitive signature of the topological phase, a feature notably missing in the trivial case (f).
		Parameters for all plots: $N=22$ sites (11 unit cells), evolution time $T=40\pi$, and recording period $\Delta t=0.4\pi$.
	}
	\label{fig::LDOS L5}
\end{figure*}

\section{LDOS IN CHIRAL SYSTEMS}

In our momentum-space lattice implementation, the expectation value of the chiral operator $\langle\Gamma(t)\rangle_{\bar{n}}$ is determined by the population imbalance between the two sublattices. For a system with $N$ momentum sites, the time-dependent chiral density can be expanded as \cite{zhang2024dynamical}
\begin{equation}
	\langle\Gamma(t/2)\rangle_{\bar{n}} = \lambda_{\bar{n}} \sum_{m=1}^{N} |c_{\bar{n}}^m|^2 e^{-i E_m t}
\end{equation}
where the coefficients $|c_{\bar{n}}^m|^2 = |\langle \psi_m |\bar{n}\rangle|^2$ with $|\psi_m\rangle$ (eigen-energy $E_m$). The local Density of States (LDOS), $\rho(\bar{n},E)$, can be directly extracted by performing a Fourier transform on $\langle\Gamma(t)\rangle_{\bar{n}}$
\begin{equation}
	 \mathcal{F}[\langle\Gamma(t/2)\rangle_{\bar{n}}] = \lambda_{\bar{n}} \sum_m |c_{\bar{n}}^m|^2 \delta(E - E_m) :=\rho(\bar{n},E)
\end{equation}
This energy-resolved spectral distribution preserves the sublattice sign information via the $\lambda_{\bar{n}}$ factor. Summing the absolute values of the LDOS across all lattice sites yields
\begin{equation}
	W(E) = \sum_{\bar{n}} |\mathcal{F}\langle\Gamma(t/2)\rangle_{\bar{n}}|
\end{equation}
which reveals the underlying energy spectrum of the SSH chain.To gain deeper insight into the spatial characteristics of the topological phase, we define the integrated spectral weight $P(\bar{n})$ by summing the LDOS over specific energy range
\begin{equation}
	P(\bar{n}) = \int_{E \in \text{range}} | \mathcal{F}\langle\Gamma(t/2)\rangle_{\bar{n}} | dE 
\end{equation}
Physically, $P(\bar{n})$ represents the average probability distribution of the wavefunctions within the chosen energy range at site $\bar{n}$. In our analysis, we partition the spectrum into three regions: $P_{\text{band,1}}$ and $P_{\text{band,2}}$, corresponding to the lower and upper bulk bands respectively, and $P_{\text{edge}}$, representing the in-gap topological states.

The numerical results presented in Fig.\ref{FIG:fig1} and Fig.\ref{fig::LDOS L5} validate the efficacy of our dynamical spectroscopy protocol for characterizing topological phases in a momentum-space Su-Schrieffer-Heeger (SSH) lattice consisting of 11 unit cells. As illustrated in the experimental scheme of Fig.\ref{FIG:fig1}(a), discrete momentum states are coupled via multi-frequency laser-driven Bragg transitions to engineer alternating tunneling amplitudes $v$ and $w$. The phase diagram in Fig.\ref{FIG:fig1}(b) defines the boundary at $v=w$, with the topological regime ($v < w$) characterized by the theoretical open-boundary-condition (OBC) energy spectrum in Fig.\ref{FIG:fig1}(c), which exhibits two degenerate zero-energy modes localized within the bandgap. This theoretical spectrum corresponds directly to the global density of states (DOS), $W(E)$, reconstructed through the Fourier transform of the chiral dynamics $\langle \Gamma(t) \rangle_{\bar{n}}$. Specifically, the sharp, isolated resonance at $E=0$ in the topological DOS of Fig.\ref{fig::LDOS L5}(c) serves as the definitive spectral signature of these edge modes , a feature notably absent in the trivial DOS of Fig.\ref{fig::LDOS L5}(f), which reveals a clear energy gap.
Furthermore, a rigorous correspondence is established between the spatial weight of the OBC eigenstates $|\psi(\bar{n})|^2$ and the integrated spectral weights $P(\bar{n})$. As shown in the topological eigenstate profile of Fig.\ref{fig::LDOS L5}(a), the zero-energy modes are strictly localized at the lattice boundaries. This is precisely reflected in the middle column of Fig.\ref{fig::LDOS L5}, where the $P_{\text{edge}}(\bar{n})$ component representing the integrated LDOS over the gap energy range peaks exclusively at the boundary sites. Conversely, the bulk state distributions $P_{\text{band},1}(\bar{n})$ and $P_{\text{band},2}(\bar{n})$, which represent the average wavefunction distributions for the lower and upper bands respectively, exhibit delocalized periodic profiles characteristic of extended bulk states across the synthetic lattice. In the trivial phase shown in Fig.\ref{fig::LDOS L5}(d)-(e), the boundary localization disappears as $P_{\text{edge}}(\bar{n})$ vanishes, and the spatial distribution is dominated by bulk-like patterns $P_{\text{bulk},1/2}(\bar{n})$ across all sites. These findings confirm that the atomic population imbalance $\langle \Gamma(t) \rangle_{\bar{n}}$ provides a direct probe of the topological edge modes in synthetic dimensions.

\section{Conclusion}

In conclusion, we have proposed a dynamical spectroscopy protocol that utilizes chiral operator dynamics to resolve the topological spectral density in momentum-space lattices. By monitoring the time evolution of sublattice-resolved atomic populations, the energy local density of states can be extracted. We further show that Fourier-transforming the chiral density can server as a direct probe of topological zero-energy edge modes. This protocol should establish a diagnostic framework for investigating topological phases in synthetic quantum matter.

\section*{Acknowledgements}
This work is supported by the National Key R$\&$D Program of China (2021YFA1401700), NSFC (Grants No. 12474267, 12074305), the Fundamental Research Funds for the Central Universities (Grant No. xtr052023002), the Shaanxi Fundamental Science Research Project for Mathematics and Physics (Grant No. 23JSZ003), Shanghai Municipal Science and Technology Major Project (Grant No.2019SHZDZX01) and the Xiaomi Young Scholar Program. We also thank the HPC platform of Xi'an Jiaotong University, where our numerical calculations was performed.

\bibliographystyle{apsrev} 
\bibliography{ref-LDOS}

\end{document}